\documentclass[a4paper,11pt]{article}
\usepackage{jinstpub} 
\usepackage{lineno}

\usepackage{physics}

\usepackage{tikz}
\usetikzlibrary{quantikz}

\title{\boldmath Ground State Preparation via Qubitization}

\author{Charles Marteau}
\affiliation{Department of Physics and Astronomy \\
University of British Columbia \\
Vancouver, BC V6T 1Z1, Canada
}

\emailAdd{cmarteau@phas.ubc.ca}

\abstract{We describe a protocol for preparing the ground state of a Hamiltonian $H$ on a quantum computer. This is done by designing a quantum algorithm that implements the imaginary time evolution operator: $e^{-\tau H}$. The method relies on the so-called ``qubitization'' procedure of Low and Chuang which, assuming the existence of a unitary encoding of the Hamiltonian $H = \bra{G} U_H \ket{G}$, produces a new operator $W_H$ whose moments are the Chebyshev polynomials of $H$ when projected on $\ket{G}$. 
Using this result and the expansion of $e^{-\tau H}$ in terms of Chebyshev polynomials we construct a circuit that implements an approximation of the imaginary time evolution operator which, at large time, projects any state on the ground state, provided a non-trivial initial overlap between the two. We illustrate our method on two models: the transverse field Ising model and a single qubit toy model.}


\makeatletter
\gdef\@fpheader{}
\makeatother

\begin{document}
\maketitle
\flushbottom

\section{Introduction}
\label{sec:intro}

Many-body quantum systems are notoriously hard to simulate on classical computers. This can be traced back to the fact that the dimension of a Hilbert space scales as the exponential of the number of degrees of freedom of the system. Quantum computers, on the other hand, should overcome this issue since they have the ability to store and apply gates directly on quantum states and are therefore expected to play a major role in the simulation of quantum systems. One of the task for which a quantum computer can achieve a significant quantum advantage is simulating the Shr\"odinger  evolution of some Hamiltonian \cite{feynman1982simulating, 
doi:10.1126/science.273.5278.1073, PhysRevLett.114.090502, Berry2015HamiltonianSW, PhysRevLett.118.010501, Low2019hamiltonian}. Another task for which quantum computers could play a major role is preparing the ground state of a many-body Hamiltonian. An efficient algorithm for ground state preparation would have numerous applications. In physics it could help understanding low-temperature quantum states such as superconductivity, superfluidity, etc. It could also help understand some intriguing properties of quantum chaos such as many-body localisation, thermalisation, etc. One can also show that any $k$-Constraint-Satisfaction-Problem can be reformulated as finding the ground state of a $k$-local Hamiltonian, drawing therefore an intimate link between a large space of classical computing tasks and the quantum task we are interested in. Efficient ground state preparations would also play a significant role in quantum chemistry and medicine where finding ground states of large molecules is necessary. Finally, there is hope to use quantum computers to simulate low-dimensional quantum gravity thanks to the gauge/gravity duality \cite{brown2021quantum}. In particular it is now understood that in this duality, the eternal black hole is non-perturbatively dual to a thermofield double state \cite{Maldacena_2003} which entangles to copies of the dual quantum mechanical system. Interestingly, even though the thermofield double is an excited state, it can also be well approximated by the ground state of an auxiliary Hamiltonian as shown in \cite{Cottrell:2018ash}. Efficient ways of preparing this ground state would therefore be of central importance in the simulation of black hole physics, especially since the gravity regime is achieved when the number of d.o.f. of the dual quantum mechanics is large.

Even though implementing this task on a quantum computer presents an advantage over classical computers, it is still an exponentially hard one \cite{TCS-066}. One can turn it into an efficiently solvable task by making the extra assumption that we have access to an oracle which prepares a state whose overlap with the ground state is under control. The intuition behind this is that, having such a state, one can then design an algorithm that will amplify its ground state component. This first step is highly non-trivial. Indeed, one can show that when picking a state $\ket{\psi}$ at random\footnote{More precisely, one can select a random state by applying a random unitary $U$ to the ground state, $\ket{\psi}=U\ket{\Omega}$, where $U$ is drawn using the Haar measure, and then compute the average value of the overlap in the ensemble: $\widebar{|\braket{\Omega}{\psi}|^2} = \int DU |\bra{\Omega}U\ket{\Omega}|^2$, which will scale like the inverse of the Hilbert space dimension.} in the Hilbert space, the typical overlap between this state and the vacuum scales as
\begin{equation}
    |\braket{\Omega}{\psi}|^2\sim 2^{-N},
\end{equation}
where $N$ is the number of qubits. One therefore overcomes this exponentially hard barrier by simply assuming that a state with controlled overlap is available.

A natural way of obtaining the ground state is by measuring the energy, using e.g. Kitaev's phase estimation algorithm \cite{Kitaev1995QuantumMA}. This will project any state on the ground state, provided the measured energy is the lowest energy.  A second class of method is based on adiabatic deformations: start with the ground state of a known Hamiltonian and slowly deform it into another Hamiltonian whose ground state is the target \cite{Jansen_2007}. This is precisely what a quantum annealer does. This method can be combined with other methods that are more computational. Indeed, one could start with some adiabatic deformation to produce a state whose overlap with the actual ground state is non trivial and then apply another algorithm which uses this state as a starting point, e.g. phase estimation \cite{PhysRevA.77.012326}. A third class of algorithms -- which has certainly the highest chance of being used on NISQ devices -- is composed of hybrid quantum-classical algorithms. One construct a parametrized quantum circuit and apply gradient descent with respect to some loss (in our case, the measured energy). An example is the variational quantum eigensolver \cite{osti_1623945} whose low depth is appealing but whose complexity is hard to estimate, both because it relies on a specific choice of ansatz for the free parameters of the circuit but also because it involves a non-convex optimization problem. One can also combine variational methods with phase estimation \cite{Parrish2019QuantumFD}. The last class of algorithms -- which will be the one our algorithm belongs to -- is expected to be relevant for fault-tolerant quantum computers. They rely on some unitary encoding of the Hamiltonian $H$ to generate a function $f(H)$ whose action (in some limit) on the state $\ket{\psi}$ is to project it on the ground state \cite{ge2018faster, Motta_2019, Lin_2020}. Example of functions are 
\begin{equation}
    \lim_{\tau\to\infty}e^{-\tau H} \quad \mathrm{or} \quad \lim_{k\to\infty}\cos^k(H).
    \label{operators}
\end{equation}
An obvious problem is that these functions are not unitary, therefore, one needs to find a way to embed them in a higher-dimensional unitary. This can be done by writing the operator as a linear combination of unitaries (LCU). Doing so, one can use a standard method \cite{Berry2015HamiltonianSW, doi:10.1137/16M1087072} to embed it using an ancilla Hilbert space whose dimension matches the number of unitary terms. As an example, one can consider the trivial unitary encoding of $H$ to be the time evolution operator $e^{-itH}$. With this operator at hand it is possible to write $\cos^k(H)$ ($k$ even) as
\begin{equation}
    \cos^k(H) =2^{-k} \sum_{\ell=-k/2}^{k/2}\begin{pmatrix}
        k \\
        k/2 + \ell
    \end{pmatrix} e^{2 i \ell H}.
\end{equation}
Using this decomposition and the LCU method one can embed $\cos^k(H)$ in a higher dimensional unitary which will require the introduction of $O(\log k)$ ancilla qubits. In this case, the unitary encoding does not require any ancilla qubits, but one can also construct unitary encodings by embedding the system's Hamiltonian as a block matrix in a higher-dimensional unitary, i.e. by constructing a unitary operator $U_H$ and an ancilla state $\ket{G}_a$ which satisfy 
\begin{equation}
   ( \bra{G}_a\otimes I_s)U_H( \ket{G}_a\otimes I_s) = H.
\end{equation}
With such an encoding, it is possible to construct unitary representations of projecting operators such as the ones in Eq. \eqref{operators}. 

In this work we propose a quantum algorithm that implements the imaginary time evolution operator assuming the existence of a unitary encoding of the Hamiltonian. The most naive strategy would be to use the power series representation of the exponential and try to turn it into a sum of unitary operators by replacing powers of $H$ with powers of the unitary encoding $U_H$. One could then use the LCU method to construct a unitary representation of the sum. Unfortunately this doesn't work since nothing guarantees that powers of $U_H$ produce powers of the Hamiltonian when projected on $\ket{G}_a$, i.e. $( \bra{G}_a\otimes I_s)U_H^n( \ket{G}_a\otimes I_s) \neq H^n$. A remedy to this issue is the so-called qubitization procedure of Low and Chuang \cite{Low2019hamiltonian}. The main idea is to use the unitary encoding $U_H$ to construct a new operator $W_H$ -- called the iterate -- whose action in the $\ket{G}_a$ direction is controlled and whose moments can be shown to produce the Chebyshev polynomials of $H$, i.e. 
$( \bra{G}_a\otimes I_s)W_H^n( \ket{G}_a\otimes I_s) = T_n(H)$. One can then use a representation of the exponential as a Chebyshev polynomial series and replace each Chebyshev polynomial by powers of the iterate to obtain a representation of the imaginary time evolution operator in terms of an infinite sum of unitaries. Finally, a truncation of this sum can be embedded in a larger unitary using the LCU method, producing therefore a unitary representation of an approximate imaginary time evolution operator.

This algorithm will require the introduction of a set of ancilla qubits whose number will scale as $\log N$ where $N$ is the truncation order of the Chebyshev series. The total number of qubits will therefore scale with the precision at which we want to implement the imaginary time evolution operator. This is to be contrasted with other types of algorithm which also involve qubitization to produce the iterate but then use it to implement certain functions $f(H)$ by means of another method called quantum signal processing \cite{PhysRevLett.118.010501, Gily_n_2019, Low2019hamiltonian}. The power of this method is that it only requires $O(1)$ number of ancillary qubits and $O\left(\sqrt{\tau}\log(1/\epsilon)\right)$ gates to implement imaginary time evolution for a time $\tau$ with error $\epsilon$. Our method turns out to be less optimal since the number of ancilla qubits scales as 
\begin{equation}
    \text{\# ancilla qubits} = O\left( \log(\tau) +  W_0\left(\frac{\tau+\log(1/\epsilon)}{\tau}\right)\right),
\label{intro1}
\end{equation}
and the number of queries, i.e. the number of control-$U_H$, control-$U_G$ (where $U_G$ is the unitary that prepares the state $\ket{G}$) and their hermitian conjugates, will scale as 
\begin{equation}
    \text{\# queries} = O\left(\left[\frac{\tau+\log(1/\epsilon)}{W_0\left(\frac{\tau+\log(1/\epsilon)}{\tau}\right)}\right]^2\right),
\label{intro2}
\end{equation}
where $W_0$ is the Lambert-W function. The sub-optimality of the number of gates compared to the quantum signal processing method will be proven in Sec. \ref{sec:scalings}.

Now even though this method is less optimal in terms of number of ancilla qubits and complexity, it is worth noting that the architecture differs quite a lot from a quantum signal processing solution. The latter would involve an alternated product of iterates and rotations whose angles need to be fine tuned for the result to match with the Chebyshev expansion of the imaginary time evolution operator, whose coefficients depend themselves on $\tau$. In practice this can be difficult. In our case, the coefficients of the of the Chebyshev expansion are identified with the (square of) the coefficients of an ancilla state which may be easier to implement. More generally, it is hard to predict what the constraints of early fault tolerant quantum computers will be, which is why having multiple architecture of quantum algorithm for the same task may turn out to be useful.

We start in Sec. \ref{sec:algorithm} by giving a detailed description of the algorithm which includes a derivation of bounds on the number of qubits and gates required for the tasks of implementing the imaginary time evolution and preparing a ground state. Then in Sec. \ref{sec:TFIM} and \ref{sec:TOY} we illustrate the method with two systems. The first one is the transverse field Ising model (TFIM) on which we run numerical experiments to test the bounds derived in the previous section. The second one is a toy model whose Hamiltonian is the Pauli-X gate. This model is simple enough that it is possible to test the algorithm with the simulated noise of a current quantum computer and obtain promising results without the use of error mitigation techniques.

\section{The Algorithm}
\label{sec:algorithm}

\subsection{Strategy}
\label{sec:strategy}

As explained in the introduction, the main character of our story will be the imaginary time evolution operator (also called the Euclidean propagator or Euclidean time evolution operator), which is obtained by analytically continuing the real-time one, i.e.
\begin{equation}
    e^{-itH}\underset{t\to -i\tau}{\to} e^{-\tau H}. 
\end{equation}
One way to see how the two differ from each other is to look at the overlap of a time-evolved state with an eigenstate of the Hamiltonian. For a state $\ket{\chi}$, the overlap on an eigenstate $\ket{\lambda}$ after some time is 
\begin{equation}
\braket{\lambda}{\chi(t)}=\braket{\lambda}{\chi}e^{-i\lambda t},
\end{equation}
if evolved with the real-time evolution operator. While, when evolved with the Euclidean one, it becomes 
\begin{equation}
\braket{\lambda}{\chi(\tau)}=\braket{\lambda}{\chi}e^{-\lambda \tau}.
\end{equation}
We observe that at large $\tau$ the overlap with the state that has the smallest eigenvalue, i.e. the ground sate, will dominate exponentially. Evolving with the Euclidean propagator therefore naturally selects the vacuum as opposed to the real-time one which preserve the absolute value of the overlap. We would like to make use of this property and design an algorithm that implements Euclidean time evolution in order to prepare ground states.

An obvious problem is that the Euclidean propagator is not unitary. A solution is to embed the system's Hilbert space $\mathcal{H}_s$ in a larger one and find a unitary operator that acts on the system factor as the Euclidean propagator. To this end we are going to assume that there exist a unitary oracle $U_H$ that acts on $\mathcal{H}_a\otimes\mathcal{H}_s$, where $\mathcal{H}_a$ is a set of ancilla qubits, and a normalized state $\ket{G}_a\in\mathcal{H}_a$ which satisfy 
\begin{equation}
    \bra{G}U_H\ket{G}=H.
\label{EncodingH}
\end{equation}
Its action on any state of the type $\ket{G}_a\otimes \ket{\psi}_s$ is 
\begin{equation}
    U_H\ket{G}\otimes \ket{\psi}=\ket{G}\otimes H\ket{\psi}+\sqrt{1-\|H\ket{\psi}\|^2}\ket{G_\psi}^\perp.
    \label{Encoding}
\end{equation}
The resulting state is a sum of two terms: the first one, aligned with $\ket{G}$, contains the action of the Hamiltonian on the system's state $\ket{\psi}$ while the second one is orthogonal to $\ket{G}$, for all values of the state $\ket{\psi}$. We have defined the state $\ket{G_\psi}^\perp$ to be normalized. Eq. \eqref{Encoding} is well defined only if $\|H\ket{\psi}\|\leq 1$ for all $\ket{\psi}$, or equivalently, if the spectral norm of $H$ is smaller than one, which is a property of all unitarily encoded Hamiltonian.\footnote{Indeed, according to Eq. \eqref{EncodingH}, for any eigenstate $\ket{\lambda}$, we have $U_H\ket{G}\ket{\lambda} = \lambda \ket{G}\ket{\lambda} + \ket{\tilde{G}_\lambda}^\perp$, where $\ket{\tilde{G}\lambda}$ is orthogonal to $\ket{G}$. Therefore $1 = \bra{G}\bra{\lambda}U^\dagger_H U_H\ket{G}\ket{\lambda} = \lambda^2 +^\perp\braket{\tilde{G}_\lambda}^\perp$, from which we deduce that $|\lambda|\leq 1.$}

For a normalized eigenstate $\ket{\lambda}$ of the Hamiltonian, Eq. \eqref{Encoding} becomes 
\begin{equation}
    U_H\ket{G}\otimes \ket{\lambda}=\lambda\ket{G_\lambda}+\sqrt{1-|\lambda|^2}\ket{G_\lambda}^\perp,
    \label{EigenEncoding}
\end{equation}
where we have defined the state $\ket{G_\lambda}_{as}\equiv\ket{G}_a\otimes \ket{\lambda}_s$. We also define $\mathcal{H}_\lambda$ to be the space spanned by the vectors $\ket{G_\lambda}$ and $\ket{G_\lambda}^\perp$. This decomposition will be useful for what follows.

Assuming the existence of such an encoding of the Hamiltonian in $\mathcal{H}_a\otimes\mathcal{H}_s$ we can now construct an algorithm that will encode the Euclidean propagator $e^{-\tau H}$. The construction relies on two steps.
\begin{itemize}
    \item First, notice that the subspace $\mathcal{H}_\lambda$ has no reason to be preserved by $U_H$, indeed the application of $U_H$ on $\ket{G_\lambda}^\perp$ can produce a vector that lies outside of $\mathcal{H}_\lambda$ and whose overlap with $\ket{G_\lambda}$ is not under control. As a consequence it is hard to predict the generic value of $\bra{G_\lambda}U^n_H\ket{G_\lambda}$, which has to be worked out case by case. One way to remedy this issue is the so-called \emph{qubitization} procedure of Low and Chuang \cite{Low2019hamiltonian}. The idea is to supplement $U_H$ with the right correction $S$ in order to ensure the preservation of $\mathcal{H_\lambda}$. The resulting operator $W_H = SU_H$ is called the \emph{iterate} and acts therefore on $\mathcal{H}_\lambda$ as an $SU(2)$ rotation. Having control over the overlap between the action of $W_H$ on $\ket{G_\lambda}^\perp$ and $\ket{G_\lambda}$, one can compute the value of $\bra{G_\lambda}W^n\ket{G_\lambda}$ which, as it turns out, is simply the $n$th Chebyshev polynomial of $\lambda$, that we denote $T_n(\lambda)$:

    \begin{equation}
    W_H^n\ket{G}\otimes \ket{\lambda}=T_n(\lambda)\ket{G_\lambda}+\sqrt{1-|T_n(\lambda)|^2}\ket{G_\lambda}^\perp.
    \label{IterateEigenEncoding}
    \end{equation}
    We'll come back to the construction of the iterate.

    \item In a second part we use the fact that $e^{-\tau H}$ admits an expansion in terms of Chebyshev polynomials: for $\|H\|\leq 1$,
    \begin{equation}
        e^{-\tau H}=J_0(i\tau)I_s+2\sum_{n=1}^\infty i^n J_n(-i\tau)T_n(-H),
        \label{ChebExp}
    \end{equation}
    where $J_n$ is the Bessel function of the first kind. To get an approximation of the Euclidean propagator we can simply truncate this sum at some order $N$. 
    
    We can construct a representation of this operator by replacing the Chebyshev polynomials by powers of the iterate. The operator becomes
    \begin{equation}
    J_0(i\tau)I_{as}+2\sum_{n=1}^{N} i^n J_n(-i\tau)(-W_H)^n,
    \label{ChebExpIterate}
    \end{equation}
    which acts on $\mathcal{H}_a\otimes\mathcal{H}_s$. We obtain a representation of the operator we wish to encode as a linear combination of unitaries (LCU). Operators of this type can be encoded in a larger Hilbert space \cite{Berry2015HamiltonianSW, doi:10.1137/16M1087072}, which we denote $\mathcal{H}_{\widebar{a}}\otimes\mathcal{H}_a\otimes\mathcal{H}_s$, where $\mathcal{H}_{\widebar{a}}$ is an ancilla Hilbert space whose dimension must be greater than the number of terms we are considering in the polynomial expansion of the propagator: $2^{n_{\widebar{a}}} \geq N + 1$.

    The end result is a unitary encoding of \eqref{ChebExpIterate}, i.e., a unitary operator $\mathbb{U}(\tau)$ that acts on $\mathcal{H}_{\widebar{a}}\otimes\mathcal{H}_a\otimes\mathcal{H}_s$ and a state $\ket{\bar{G}}$ in $\mathcal{H}_{\widebar{a}}$ which satisfy
    \begin{equation}
    _{\widebar{a}}\bra{\bar{G}}\mathbb{U}(\tau) \ket{\bar{G}}_{\widebar{a}} = J_0(i\tau)I_{as}+2\sum_{n=1}^{N} i^n J_n(-i\tau)(-W_H)^n.
    \end{equation}
    By construction this master operator is also a unitary encoding of an approximation of the Euclidean propagator since we can further project it on the state $\ket{G}_a$ to turn the powers of the iterate into Chebyshev polynomials of $H$: 
    \begin{equation}
    \left(\bra{\bar{G}}\otimes \bra{G}\right)\mathbb{U}(\tau)\left( \ket{\bar{G}}\otimes \ket{G}\right) = J_0(i\tau)I_s+2\sum_{n=1}^N i^n J_n(-i\tau)T_n(-H) \simeq e^{-\tau H}.
    \end{equation}
    This concludes our description of the algorithm's architecture. We now give more details on the various steps.
\end{itemize}

\subsection{Qubitization} 
\label{sec:qubitization}

As mentioned previously the unitary encoding $U_H$ has no reason to preserve the subspace $\mathcal{H}_\lambda$. Clearly $U_H\ket{G_\lambda}$ belongs to it but we have no control over the action of $U_H$ on $\ket{G_\lambda}^\perp$. What we are looking for is an operator $W_H = SU_H$ that acts like $U_H$ on $\ket{G_\lambda}$ and who acts on $\mathcal{H}_\lambda$ as an $SU(2)$ rotation, 
\begin{equation}
    (W_H)_{\mathcal{H}_\lambda} =
    \begin{pmatrix}
    \lambda && -\sqrt{1-\vert \lambda \vert^2}\\
    \sqrt{1-\vert \lambda \vert^2} && \lambda
    \end{pmatrix}.
    \label{MatrixElements}
\end{equation}
The authors of \cite{Low2019hamiltonian} show that even in the case where no such correction $S$ exists for $U_H$, it is always possible to construct a quantum circuit $\tilde{U}_H$ with one more qubit for which a canonical correction $\tilde{S}$ exists. The constructions relies on having access to controlled-$U_H$ and controlled-$U^\dagger_H$ and the new operator is 
\begin{equation}
    \tilde{U}_H = \quad
    \begin{quantikz}
     \qw   & \octrl{1}  & \ctrl{1} &  \qw 
    \\
     \qw  & \gate{U_H} & \gate{U^\dagger_H} & \qw 
    \end{quantikz}
    \quad.
\end{equation}
One can easily show that it unitarily encodes the same Hamiltonian when projected on the state $\ket{\tilde{G}} = \frac{1}{\sqrt{2}}(\ket{0}+\ket{1})\otimes \ket{G}$. For this new operator they show that the canonical correction is the product of a reflection w.r.t the state $\ket{\tilde{G}}$ and the SWAP operator of the new qubit,
\begin{equation}
    \tilde{S} = \left[\left(2\ketbra{\tilde{G}}{\tilde{G}}- I_a\right)\otimes I_s\right]\left[\left(\ketbra{0}{1}+\ketbra{1}{0}\right)\otimes I_{as}\right].
\end{equation}
The latter is such that the the matrix elements of the operator $\tilde{W}_H = \tilde{S}\tilde{U}_H$, when projected on the eigen-subspaces match Eq. \eqref{MatrixElements}. Since this construction is completely generic and does not depend on the details of the unitary encoding given by $U_H$ and $\ket{G}$ we can assume without loss of generality that the correction $S$ exists and that the $SU(2)$ encoding $W_H$ can be constructed.

In \cite{Low2019hamiltonian}, it is also shown that if the unitary encoding satisfies the constraint $U^2_H = I_{as}$ then there is no need for an additional qubit and the correction is simply given by the reflection w.r.t to the ancilla state $\ket{G}$, 
\begin{equation}
    S = (2\ketbra{G}{G}- I_a)\otimes I_s.
\end{equation}
We will see that the TFIM falls into that category, which will simplify the implementation. 

The power of this construction is that we have now full control over the action of any power of the iterate $W_H$ on states that are parallel to $\ket{G}$. Indeed it is easy to show that, provided Eq. \eqref{MatrixElements} is satisfied,  repeated applications of the iterate on $\ket{G_\lambda} = \ket{G}\otimes \ket{\lambda}$ will produce the Chebyshev polynomials of $\lambda$ at the corresponding order:
\begin{equation}
    W_H^n\ket{G}\otimes \ket{\lambda}=T_n(\lambda)\ket{G_\lambda}+\sqrt{1-|T_n(\lambda)|^2}\ket{G_\lambda}^\perp.
\end{equation}
Now since this is true for any eigenstate of $H$ and that all the states $\ket{G_\lambda}^\perp$ are perpendicular to $\ket{G}$, we deduce that $W_H^n$ is a unitary encoding of the corresponding Chebyshev polynomial of $H$: $\bra{G}W_H^n\ket{G} = T_n(H)$.

\subsection{Encoding of the Chebyshev expansion}
\label{sec:encoding}

Armed with this implementation of the Chebyshev polynomials of $H$ we would like to construct a unitary encoding of a truncated polynomial expansion of the Euclidean propagator. We will therefore make use of the formula \eqref{ChebExp} which, when truncated, gives an approximation of the Euclidean propagator. We can turn the sum into a sum of unitary operators by replacing each Chebyshev polynomial of $H$ by a power of the iterate $W_H$. The new operator is now a linear combination of unitary operators whose coefficients are $\{J_0(i\tau), 2i^nJ_n(-i\tau)\}$ with corresponding operators $\{I, (-W_H)^n\}$. There are $N + 1$ of them which corresponds to the number of terms in the expansion. 

A standard way to construct a unitary representation of a linear combination of unitary operators is to introduce a new set of ancilla qubits $\mathcal{H}_{\widebar{a}}$ in which we are going to encode the coefficients of the sum. We therefore introduce $n_{\widebar{a}}$ new ancilla qubits where $n_{\widebar{a}}$ is the smallest integer such that $2^{n_{\widebar{a}}} \geq N + 1$, i.e. such that the dimension of $\mathcal{H}_{\widebar{a}}$ is sufficiently big to accomodate all the coefficients. We then construct the state\footnote{One can check that the coefficients of the Chebyshev expansion are all real.}
\begin{equation}
    \ket{\bar{G}}_{\widebar{a}} = \frac{1}{J_0(i\tau)+\sum_{n = 1}^{N}2i^nJ_n(-i\tau)}\left(\sqrt{J_0(i\tau)}\ket{0}_{\widebar{a}}+ \sum_{n = 1}^{N}\sqrt{2i^nJ_n(-i\tau)}\ket{n}_{\widebar{a}}\right).
    \label{GbarState}
\end{equation}
We then assume that we have access to a controlled version of the iterate and construct the following operator,
\begin{equation}
    \mathbb{U}(\tau) =  \sum_{n = 0}^{N}\ketbra{n}{n}_{\widebar{a}}\otimes (-W_H)^n.
\end{equation}
This operator is unitary and trivially encodes the truncated version of the expansion given in Eq. \eqref{ChebExpIterate} when projected on the state $\ket{\bar{G}}$.

We can now bring all the pieces together: the master operator $\mathbb{U}(\tau)$ encodes the truncated power expansion which involves the iterate,
\begin{equation}
\begin{split}
    \mathbb{U}(\tau)\ket{\bar{G}}\ket{G}\ket{\psi} &= \ket{\bar{G}}\left(J_0(i\tau)I_{as}+2\sum_{n=1}^{N} i^n J_n(-i\tau)(-W_H)^n\right)\ket{G}\ket{\psi}+\ket{\bar{G}_\psi}^\perp\\
    &=\ket{\bar{G}}\ket{G}\left(J_0(i\tau)I_s+2\sum_{n=1}^{N} i^n J_n(-i\tau)T_n(-H)\right)\ket{\psi}+\ket{G_\psi}^\perp+\ket{\bar{G}_\psi}^\perp\\
    &\simeq\ket{\bar{G}}\ket{G}e^{-\tau H}\ket{\psi}+\ket{G_\psi}^\perp+\ket{\bar{G}_\psi}^\perp.
\end{split}
\end{equation}
To recover the system's state we simply project the full state on the ancilla state $\ket{\bar{G}}_{\widebar{a}}\otimes\ket{G}_a$ since we know that the remainder is orthogonal by construction. In practice this product state is prepared by two unitary gates $U_{\bar{G}}$ and $U_G$ that acts on a reference state $\ket{0}_{\widebar{a}}\otimes \ket{0}_a$, we therefore first uncompute this state, i.e. we apply the inverse gates $U^{\dagger}_{\bar{G}}$ and $U^{\dagger}_G$, and then project on the reference state. Finally, to increase the probability of success when projecting on the reference state, we can supplement the algorithm with runs of amplitude amplification. 

The architecture of the algorithm can be summarized by the following quantum circuit:
\begin{equation}
    \begin{quantikz}
        \lstick{$\ket{\bar{G}}_{\widebar{a}}$} & \ctrl{1} & \ctrl{1} & \ctrl{1} & \ctrl{1} & \qw \rstick[wires=3]{$\ket{\bar{G}}\ket{G}e^{-\tau H}\ket{\psi}+\ket{G_\psi}^\perp+\ket{\bar{G}_\psi}^\perp$} 
        \\
        \lstick{$\ket{G}_a$} & \gate[wires=2]{(-W_H)^0} & \gate[wires=2]{(-W_H)^1} & \gate[wires=2]{\ldots} &
        \gate[wires=2]{(-W_H)^{N}} &
        \qw
        \\ 
        \lstick{$\ket{\psi}_s$} & & & & &\qw
    \end{quantikz}
    ,
    \label{CircuitTotal}
\end{equation}
where each block corresponds to a controlled version of a power of the iterate. More precisely, the $n$th power of the iterate is executed when the state of $\mathcal{H}_{\widebar{a}}$ is $\ket{n}_{\widebar{a}}$.

\subsection{Scalings}
\label{sec:scalings}

\paragraph{Minimal distance} We have now an algorithm that implements the Euclidean propagator on the system factor of the Hilbert space. Changing the value of the Euclidean time $\tau$ amounts to a change of the coefficients of the state $\ket{\bar{G}}$, given in Eq. \eqref{GbarState}. In principle, when increasing $\tau$, the resulting state should get increasingly closer to the actual ground state $\ket{\Omega}$ of the Hamiltonian. This is of course not exactly true since we are actually implementing a truncation of the Euclidean propagator. After a certain critical time $\tau_*$ the error due to the truncation will start to become relevant and the distance will now grow. This means that there is a \emph{minimal} non-vanishing achievable distance between the simulated state and the ground state.

Let us propose a model for the evaluation of this critical time. We define the truncation error $R$ in the following way,
\begin{equation}
    \begin{split}
        e^{-\tau H} &=J_0(i\tau)I_s+2\sum_{n=1}^{N} i^n J_n(-i\tau)T_n(-H)+R(\tau),\\
        R(\tau) &= 2\sum_{n=N+1}^{\infty} i^n J_n(-i\tau)T_n(-H).
    \end{split}
\end{equation}
With this definition, if we feed the algorithm with an arbitrary system state $\ket{\psi}$, the latter will be projected on the state
\begin{equation}
    \ket{\psi(\tau)} = \frac{\left(U(\tau) - R(\tau)\right)\ket{\psi}}{\norm{\left(U(\tau) - R(\tau)\right)\ket{\psi}}},
    \label{TruncatedEvolvedState}
\end{equation}
where $U(\tau)$ denotes the ideal Euclidean propagator. Assuming that the truncation error is small, i.e. that powers of $R$ are much smaller than $R$ itself, the projection of this state on the vacuum $\ket{\Omega}$ becomes,
\begin{equation}
    \braket{\Omega}{\psi(\tau)}=\frac{\bra{\Omega}U(\tau)\ket{\psi}}{\norm{U(\tau)\ket{\psi}}} - \frac{\bra{\Omega}R(\tau)\ket{\psi}}{\norm{U(\tau)\ket{\psi}}} +
    \frac{\bra{\Omega}U(\tau)\ket{\psi}}{\norm{U(\tau)\ket{\psi}}^3}
    \bra{\psi}U(\tau)R(\tau)\ket{\psi} + O(R^2).
    \label{Approx1}
\end{equation}
The first term corresponds to the state we would obtain if the Euclidean propagation was implemented perfectly. If we define $\Delta E = E_{\mathrm{exc}}-E_\Omega$ to be the energy difference between the first excited state $\ket{\mathrm{exc}}$ that has a non-trivial overlap with $\ket{\psi}$ and the ground state energy, then, for $\tau$ sufficiently greater than the characteristic time $1/\Delta E$, the first term becomes
\begin{equation}
    \frac{\bra{\Omega}U(\tau)\ket{\psi}}{\norm{U(\tau)\ket{\psi}}} \sim e^{i\theta_\Omega}\left(1-\frac{1}{2}\,\frac{\psi_{\mathrm{exc}}^2}{\psi_\Omega^2}e^{-2\tau\Delta E}\right),
    \label{VacuumApprox}
\end{equation}
where we have defined the projection of the system state $\ket{\psi}$ on the ground state and $\ket{\mathrm{exc}}$ to be: $\braket{\Omega}{\psi} = \psi_\Omega e^{i\theta_\Omega}$ and $\braket{\mathrm{exc}}{\psi} = \psi_\mathrm{exc} e^{i\theta_\mathrm{exc}}$. The distance between the ideal time evolved state and $\ket{\Omega}$ is therefore measured by $\varepsilon(
\tau) \equiv \frac{1}{2}\frac{\psi_{\mathrm{exc}}^2}{\psi_\Omega^2}e^{-2\tau\Delta E}$.

\begin{figure}[htbp]
\centering
\includegraphics[width=.7\textwidth]{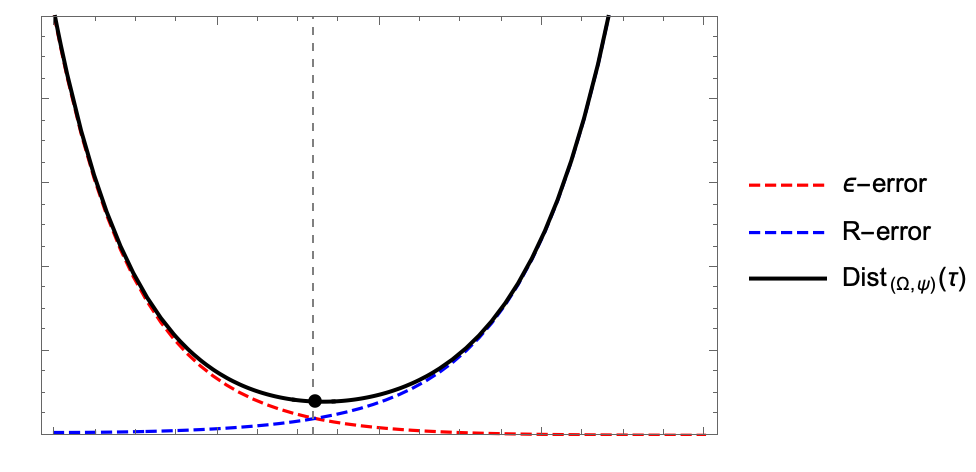}
\caption{Schematic plot of the distance between $\ket{\psi(\tau)}$ and $\ket{\Omega}$. The solid curve is $D_{\Omega,\psi(\tau)}$, the red dashed curve is the ideal distance (no truncation) and the blue dashed curve is the truncation error. At first the Euclidean propagator is well approximated by the truncated sum, the distance therefore decreases exponentially and is well approximated by the first term in Eq. \eqref{Distance}. After a critical time $\tau_*$ the distance increases again, the truncation error -- i.e. the second term in Eq. \eqref{Distance} -- start to dominate.}
\label{fig:Distance}
\end{figure}

For times sufficiently large compared to $1/\Delta E$, $\varepsilon$ becomes small and, since we are also in a regime where $R$ is small, we can further approximate the result by truncating it at a certain order in both $R$, $\varepsilon$ and their products. From Eq. \eqref{VacuumApprox} we deduce
\begin{equation}
    \frac{U(\tau)\ket{\psi}}{\norm{U(\tau)\ket{\psi}}}\sim e^{i\theta_\Omega}\left(1-\varepsilon(\tau)\right)\ket{\Omega}+\sqrt{\,2\,\varepsilon(\tau)}\ket{\chi},
\end{equation}
where $\ket{\chi}$ is a normalized state orthogonal to $\ket{\Omega}$. We plug this value in Eq. \eqref{Approx1} and expand at order $3/2$ in $\varepsilon$ and $R$ to obtain
\begin{equation}
   D_{\Omega,\psi(\tau)} \equiv 1 - |\braket{\Omega}{\psi(\tau)}|^2\sim 
    2\left(\varepsilon(\tau) -
    \sqrt{\,2\,\varepsilon(\tau)}
    \,\frac{\mathrm{Re}\bra{\chi}R(\tau)\ket{\psi}}{\norm{U(\tau)\ket{\psi}}}
    \right).
    \label{Distance}
\end{equation}
We observe the distance between the ground state and the Euclidean time evolved state receives two contributions. The first one is proportional to $\varepsilon(\tau)$ and decays exponentially, it corresponds to the decreasing distance between the ground state and the state that would have been evolved with the ideal Euclidean propagator $e^{-\tau H}$. The second one is proportional to $R(\tau)$ and, as we will see, is exponentially growing at late times. It is the error due to the truncation of the Chebyshev expansion. Therefore the time at which the distance is minimal should correspond approximately to the time at which the two terms in Eq. \eqref{Distance} are of the same magnitude, see Fig. \ref{fig:Distance}.

Let us now focus on the truncation error term which involves the quantity $\mathrm{Re}\bra{\chi}R(\tau)\ket{\psi}$. We would like to know how this quantity scales with $\tau$ and the the truncation order $N$. We have that 
\begin{equation}
    |\mathrm{Re}\bra{\chi}R(\tau)\ket{\psi}| \leq |\bra{\chi}R(\tau)\ket{\psi}|.
\end{equation}
Now the term on the right hand side is the norm of an infinite sum of terms that can be bounded by the sum of the norms,
\begin{equation}
  |\mathrm{Re}\bra{\chi}R(\tau)\ket{\psi}| \leq
  2\sum_{n=N+1}^{\infty} i^n J_n(-i\tau)|\bra{\chi}T_n(-H)\ket{\psi}|.
\end{equation}
This inequality is true because the coefficients $i^n J_n(-i\tau)$ are all positive and real for $\tau \geq 0$. Moreover, the Chebyshev polynomials have the property that their absolute value is smaller than one on the interval $[-1,1]$, therefore, since the norm of $H$ is smaller than one, we have that
\begin{equation}
  |\mathrm{Re}\bra{\chi}R(\tau)\ket{\psi}| \leq
  2\sum_{n=N+1}^{\infty} i^n J_n(-i\tau).
\end{equation}
We can use the expression of the Bessel function as a infinite sum to obtain 
\begin{equation}
  |\mathrm{Re}\bra{\chi}R(\tau)\ket{\psi}| \leq
  2\sum_{m=0}^{+\infty}\frac{1}{m!}\left(\frac{\tau}{2}\right)^m\sum_{n=N+1}^{+\infty}\frac{1}{(m+n)!}\left(\frac{\tau}{2}\right)^{n+m}.
\end{equation}
We observe that the right hand side involves a remainder of the Taylor series of $e^{\tau/2}$. The latter can be bounded in the following way,
\begin{equation}
    \begin{split}
    \sum_{n=N+1}^{+\infty}\frac{1}{(m+n)!}\left(\frac{\tau}{2}\right)^{n+m} 
    &=\frac{\left(\frac{\tau}{2}\right)^{m+N+1}}{(m+N+1)!}\sum_{n=0}^{+\infty}\frac{1}{(m+N+2)(m+N+3)\ldots(m+N+1+n)}\left(\frac{\tau}{2}\right)^n\\
    &\leq \frac{\left(\frac{\tau}{2}\right)^{m+N+1}}{(m+N+1)!}\sum_{n=0}^{+\infty}\frac{1}{n!}\left(\frac{\tau}{2}\right)^n=\frac{\left(\frac{\tau}{2}\right)^{m+N+1}}{(m+N+1)!}\,e^{\tau/2}.
    \end{split}
\end{equation}
Which implies the following inequality
\begin{equation}
  |\mathrm{Re}\bra{\chi}R(\tau)\ket{\psi}| \leq
  2\sum_{m=0}^{+\infty}\left(\frac{\tau}{2}\right)^{2m+N+1}\frac{e^{\tau/2}}{m!(m+N+1)!}=2\, e^{\tau/2}\,I_{N+1}(\tau).
  \label{RBound}
\end{equation}
Using this upper bound for $|\mathrm{Re}\bra{\chi}R(\tau)\ket{\psi}|$ we can find a \emph{lower} bound for $\tau_*$. Indeed, we have found that the blue dashed curve of Fig. \ref{fig:Distance} lies under some monotonically increasing function. The latter will intersect the $\varepsilon$-error curve (red dashed) at a time below $\tau_*$.  This lower bound, denoted $\bar{\tau}_*$, is therefore obtained by equating the absolute values of the two terms that are contributing to the distance (see Eq. \eqref{Distance}) and replacing $|\mathrm{Re}\bra{\chi}R(\tau)\ket{\psi}|$ by the bound that we have just derived. Moreover, if we approximate $\norm{U(\widebar{\tau}_*)\ket{\psi}}$ by its largest contribution, i.e. the one coming from the ground state, $\norm{U(\widebar{\tau}_*)\ket{\psi}}\sim \psi_\Omega e^{-\widebar{\tau}_* E_\Omega}$, one obtains\footnote{It should be kept in mind that we assume that the spectral norm of our Hamiltonian is smaller than one, therefore $E_{\mathrm{exc}}\in [-1,1]$. This also means that Eq. \eqref{ToSolve} admits a solution for any allowed values of $E_{\mathrm{exc}}$ and $\psi_{\mathrm{exc}}$.}
\begin{equation}
    \frac{\psi_{\mathrm{exc}}}{4}e^{-\widebar{\tau}_*\left(E_{\mathrm{exc}}+\frac{1}{2}\right)}=I_{N+1}(\widebar{\tau}_*).
    \label{ToSolve}
\end{equation}
The left hand side is an exponentially decreasing function while the right hand side initially grows as a power of $\tau$ and then exponentially. We approximate the Bessel function by its large-$n$ behaviour: $I_n(\tau)\underset{n\sim +\infty}{\sim}\frac{1}{\sqrt{2\pi n}}\left(\frac{e\tau}{2n}\right)^n$. With this replacement we obtain a solution for $\widebar{\tau}_*$
\begin{equation}
    \tau_*\geq\widebar{\tau}_* = \frac{N+1}{E_{\mathrm{exc}}+\frac{1}{2}}\,W_0\left(2\,e^{-1}\left(E_{\mathrm{exc}}+\frac{1}{2}\right)\left(\frac{\psi_{\mathrm{exc}}}{4}\,\sqrt{2\pi (N+1)}\right)^{1/(N+1)}\right),
\end{equation}
where $W_0$ is the Lambert-$W$ function. This bound, as a function of $N$, is monotonically increasing, at large $N$ it grows linearly with a slope $\frac{W_0\left(2\,e^{-1}\left(E_{\mathrm{exc}}+\frac{1}{2}\right)\right)}{E_{\mathrm{exc}}+\frac{1}{2}}$. This is consistent with the qualitative observation that for larger truncating order $N$, the truncated Chebyshev expansion will be a good approximation of the Euclidean propagator for larger times. Interestingly, the (large $N$) asymptotic bound does not depend on the initial state.

\begin{figure}[htbp]
\centering
\includegraphics[width=.8\textwidth]{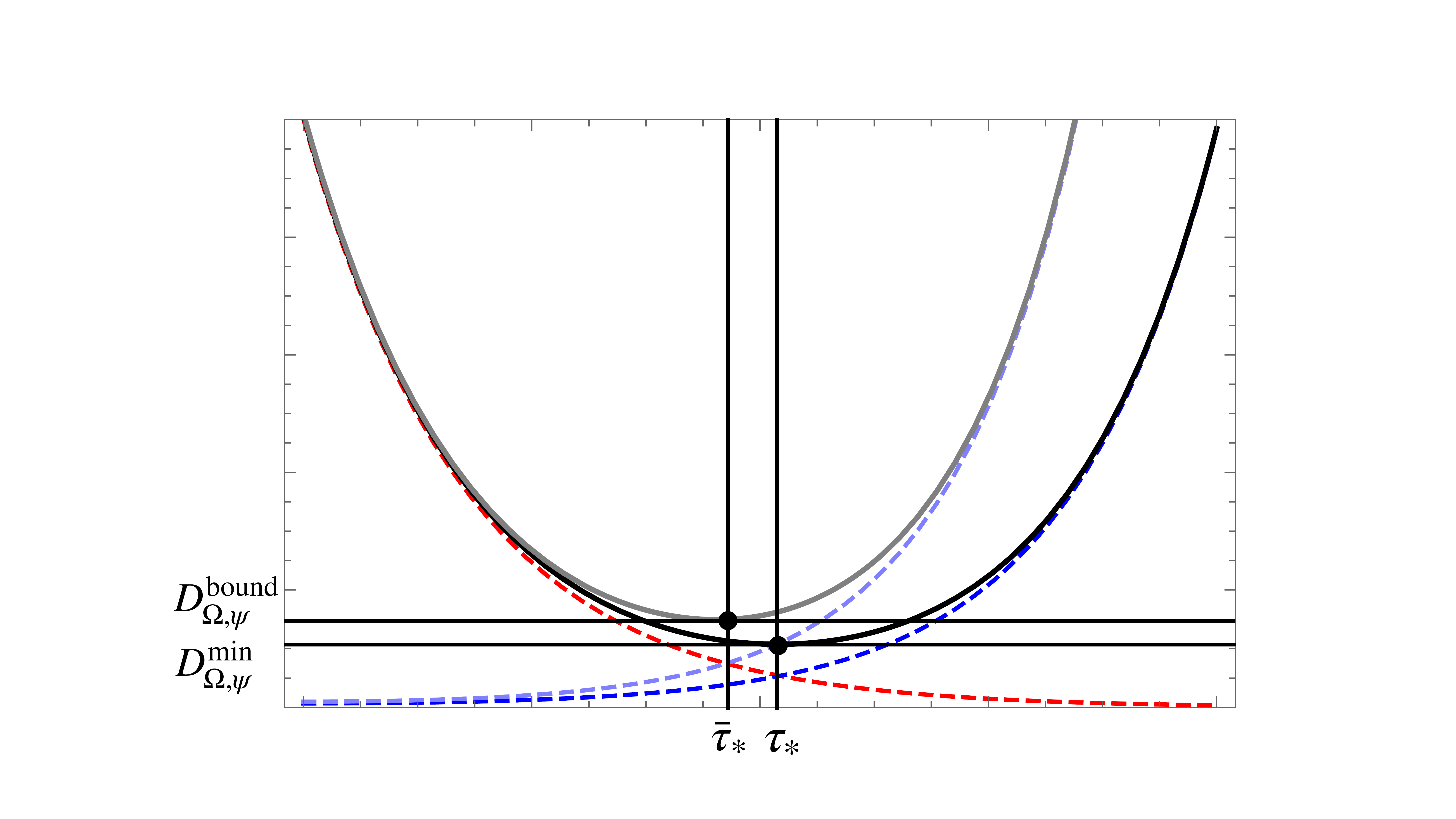}
\caption{Illustration of the method to derive the bound on the minimal distance. The solid black line corresponds to the actual distance between $\ket{\psi(\tau)}$ and the ground state $\ket{\Omega}$ while the gray one is the upper bound of Eq. \eqref{DistanceBound}. The blue dashed curve is the truncation error and the light blue one is the upper bound on the truncation error. We observe that by finding the time at which the light blue dashed curve intersects the ideal distance (red dashed curve) and then evaluating the bound on the distance at that time, we obtain an upper bound on the minimal distance.}
\label{fig:DistanceBound}
\end{figure}

We can use those results to compute a bound on the minimal distance achievable. Indeed using Eq. \eqref{Distance} we deduce that the distance is bounded by 
\begin{equation}
\begin{split}
    D_{\Omega,\psi(\tau)}
    &\leq
    2\,\varepsilon(\tau)\left(1 +
    \sqrt{\,2}
    \,\frac{|\mathrm{Re}\bra{\chi}R(\tau)\ket{\psi}|}{\sqrt{\,\varepsilon(\tau)}\norm{U(\tau)\ket{\psi}}}
    \right),\\
    &\leq 
    2\,\varepsilon(\tau)\left(1 +
    \sqrt{\,2}
    \,\frac{2\, e^{\tau/2}\,I_{N+1}(\tau)}{\sqrt{\,\varepsilon(\tau)}\norm{U(\tau)\ket{\psi}}}
    \right).
    \end{split}
\label{DistanceBound}
\end{equation}
Where the second inequality follows from Eq. \eqref{RBound}. To learn about the minimal achievable distance we will evaluate this bound at time $\tau = \bar{\tau}_*$. Indeed we know that the distance is minimal at $\tau = \tau_*$ but we only have a lower bound for $\tau_*$. Therefore by evaluating the bound of Eq. \eqref{DistanceBound} at $\tau =\bar{\tau}_* $ we get an \emph{upper} bound on the minimal distance, see Fig. \ref{fig:DistanceBound}. 

Moreover we know that the time $\widebar{\tau}_*$ is defined to be the time at which the second term of Eq. \eqref{DistanceBound} is very close to one. Which means that 
\begin{equation}
    D^{\mathrm{min}}_{\Omega,\psi(\tau)}\leq 4\,\varepsilon(\widebar{\tau}_*).
\end{equation}
For $N$ sufficiently large we can replace the bound $\widebar{\tau}_*$ by its linear approximation as a function of the truncation order $N$. We obtain\footnote{One could think that the bound is irregular when $E_\mathrm{exc} = -1/2$, which is not forbidden by our assumption that the spectral norm of $H$ is smaller than one. Actually one can check that the function $\frac{W_0(2e^{-1}x)}{x}$ is perfectly regular on the domain $(-1/2, +\infty)$. It is a monotonically decreasing function which satisfies $\underset{x\to -1/2}{\lim} = 2$ and $\underset{x\to +\infty}{\lim} = 0$, which means that $E_\mathrm{exc}$ can be arbitrarily small to $-1$.}
\begin{equation}
    D^{\mathrm{min}}_{\Omega,\psi(\tau)}\leq 2\,\frac{\psi_{\mathrm{exc}}^2}{\psi_\Omega^2}\,\exp\left[-\frac{2\,W_0\left(2\,e^{-1}\left(E_{\mathrm{exc}}+\frac{1}{2}\right)\right)\Delta E}{E_{\mathrm{exc}}+\frac{1}{2}}\,N\right],
\end{equation}
from which we read that the upper bound on the minimal distance decreases exponentially as a function of $N$. This qualitative behavior is expected since increasing $N$ increases the time during which the algorithm well approximates the actual Euclidean propagator, leading to a better approximate projection on the ground state. Conversely the number of terms in the Chebyshev expansion necessary to achieve an error $\epsilon$ between the state produced by the algorithm and the ground state scales like $\log(1/\epsilon)$. 

Finally we can deduce a scaling for the number of queries necessary to achieve an error $\epsilon$. Indeed, the number of queries to the unitary encoding of the Hamiltonian $U_H$ and the unitary $U_G$ that creates the state $\ket{G}$ are of order $O\left(N^2\right)$. This is because the algorithm given in Eq. \eqref{CircuitTotal} implements sequentially all the $n$th powers of $W_H$. Since each implementation of $W_H$ counts for at most two gates $U_H$ and two gates $U_G$ (see Sec. \ref{sec:qubitization}), the total numbers of $U_H$ and $U_G$ gates are
\begin{equation}
\mathrm{\#}\,U_H = N(N+1), \quad \mathrm{\#}\,U_G = N(N+1)+1, 
\end{equation}
where the $+1$ in $\mathrm{\#}\,U_G$ comes from the additional gate that initializes the state $\ket{G}_a$. Now to achieve an error $\epsilon$ we need $\log\left(1/\epsilon\right)$, therefore the same error is achieved with $O\left(\log^2\left(1/\epsilon\right)\right)$ gates.

\paragraph{Total number of qubits} We would like to estimate the number of qubits necessary to execute the algorithm. There are three types of qubits: the system's qubits $\mathcal{H}_s$, a first family of ancilla qubits $\mathcal{H}_a$ for the unitary encoding of the system's Hamiltonian and a second family of ancilla qubits $\mathcal{H}_{\widebar{a}}$ for the unitary encoding of the Chebyshev expansion. In practice one would also introduce other ancilla qubits for transpiling the circuit but we will not consider them here. 

The number of system qubits $n_s$ scales as $\log d$, were $d$ is the dimension of $\mathcal{H}_s$. For the number of ancilla qubits $n_a$ we know that the $d^2$ Pauli matrices form a unitary basis for $d\times d$ matrices and since $2^{n_a}$ should scale like the number of terms in the unitary decomposition of $H$ we conclude that $n_a$ scales at most like $\log d$ too. Finally the number of ancilla qubits $n_{\widebar{a}}$ is such that $2^{n_{\widebar{a}}}$ scales like the number of terms in the Chebyshev expansion, therefore $n_{\widebar{a}}\sim \log N$. We conclude 
\begin{equation}
    \#\mathrm{qubits} \sim \log d + \log N.
\end{equation}
Now we know that the minimal truncation order $N$ to achieve an error $\epsilon$ between the simulated state and the ground state is $N \sim \log(1/\epsilon)$. We therefore conclude that the total number of qubits to achieve an error $\epsilon$ is 
\begin{equation}
    \#\mathrm{qubits} \sim \log d + \log \log(1/\epsilon).
\end{equation}

\paragraph{Euclidean propagator approximation} We would like to end this section by asking another question: what is the query complexity necessary for this algorithm to implement the Euclidean propagator with error $\epsilon$. A reasoning similar to the one exposed in the previous paragraph leads to the bound
\begin{equation}
    |R(\tau)|\leq 2\,e^{\tau/2}I_{N+1}(\tau),
\end{equation}
which is a bound on the error between the ideal propagator and the truncated one. Therefore, by inverting this equation we can obtain the truncation order $N$ which will guarantee an error smaller than $\epsilon$. We seek to solve the equation 
\begin{equation}
2\,e^{\tau/2}I_{N+1}(\tau) = \epsilon,
\end{equation}
where $N$ is the unknown. For an error $\epsilon$ sufficiently small and therefore $N$ sufficiently large we can approximate the Bessel function by its large-$n$ behaviour: $I_n(\tau)\underset{n\sim +\infty}{\sim}\frac{1}{\sqrt{2\pi n}}\left(\frac{e\tau}{2n}\right)^n$. With this replacement and taking the $\log$ we obtain the equation 
\begin{equation}
    \log(1/\epsilon)+\tau/2 = N\log N - N\log\tau,
\end{equation}
where we have made the approximation $N+\mathrm{cst}\sim N$, $\tau+\mathrm{cst}\sim \tau$ and similarly for their logs. This equation is solved by 
\begin{equation}
    N = \tau\, \exp \left[W_0\left(\frac{\tau/2+\log(1/\epsilon)}{\tau}\right)\right]=\frac{\tau/2+\log(1/\epsilon)}{W_0\left(\frac{\tau/2+\log(1/\epsilon)}{\tau}\right)}.
    \label{ScalingTruncation}
\end{equation}
As observed at the end of the paragraph on the minimal distance, our algorithm requires $O\left(N^2\right)$ queries, which leads to the formula \eqref{intro2} advertised in the introduction for the number of queries. Moreover, it was also observed in the previous paragraph that the number of ancilla qubits scales as $\log N$, which leads to the other formula \eqref{intro1} advertised in the introduction for the number of ancilla qubits.

We can first look at the large-$\tau$ limit of Eq. \eqref{ScalingTruncation} with fixed error $\epsilon$. In that case the number of queries scales as
\begin{equation}
    \#\mathrm{queries}\underset{\tau\sim +\infty, \, \epsilon \,\mathrm{fixed}}{\sim}\, \tau^2.
\end{equation}
Now, if we fix $\tau$ and take the small-$\epsilon$ limit, we can use the asymptotic form of the Lambert $W$ function $W_0(x)\underset{x\sim+\infty}{\sim}\log x-\log\log x+o(1)$ to obtain that the number of queries scales as 
\begin{equation}
    \#\mathrm{queries}\underset{\epsilon\sim 0, \, \tau \,\mathrm{fixed}}{\sim}\,\frac{\log^2(1/\epsilon)}{\log^2\left(\tau^{-1}\log(1/\epsilon)\right)}.
\end{equation}
It is easy to check that this function grows faster than $\log(1/\epsilon)$. This implies that the scaling for the number of queries is sub-optimal as compared to the $O(\sqrt{\tau}\log(1/\epsilon))$ obtained using quantum signal processing \cite{Gily_n_2019}, both as a function of $\tau$ and $\epsilon$. However, it should be noted that the gain with quantum signal processing versus our method is polynomial as a function of $\tau$ and sub-polynomial as a function of $\log(1/\epsilon)$.

\section{Transverse field Ising model}
\label{sec:TFIM}

We would like to illustrate our method with a well known quantum system: the periodic TFIM. We start by constructing a unitary encoding of the TFIM Hamiltonian. This unitary encoding $U_H$, as we will prove, satisfies the identity $U_H^2 = I_{as}$, which simplifies the qubitization procedure (see Sec. \ref{sec:qubitization}). Details on the practical construction of the algorithm are given together with estimations of the number of qubits and gates. We conclude this section with numerical simulations of the two-sites TFIM on which we test the validity of the bounds derived in Sec. \ref{sec:scalings}.

\paragraph{Unitary encoding of the Hamiltonian}

The Hamiltonian of the TFIM is 
\begin{equation}
    H= \sum_{i=1}^L X_iX_{i+1}+g\sum_{i=1}^LZ_i,
\end{equation}
where $L$ is the number of sites, $g$ the coupling to the transverse field and $\{X_i, Y_i, Z_i\}$ the Pauli matrices at site $i$. The sites are periodic, i.e., $i+L=i$. The system's Hilbert space $\mathcal{H}_s$ is that of $L$ qubits, therefore $\mathrm{dim}(\mathcal{H}_s)= 2^L$. 

This Hamiltonian is a linear combination of unitary operators since the Pauli matrices are Hermitian and square to the identity. We would like to use this property of the TFIM to find a unitary encoding. For simplicity we will choose $L$ such that $\exists\, n_a\in \mathbb{N}$ which satisfies $2^{n_a}=2L$, which corresponds to the number of terms in the Hamiltonian. We therefore introduce $n_a$ ancilla qubits whose Hibert space will be the host of the state\footnote{The multi-qubit state $\ket{i}$ is identified with the state $\ket{q_0\ldots q_{n_a-1}}$ such that $\sum_{k=0}^{n_a-1}q_k 2^k = i$.}
\begin{equation}
    \ket{G}=\frac{\sum_{i=0}^{L-1}\ket{i}+\sqrt{g}\sum_{i=L}^{2L-1}\ket{i}}{\sqrt{L(1+|g|)}}.
\end{equation}
Consider also the operator
\begin{equation}
    U_H = \sum_{i=0}^{L-1} \ketbra{i}\otimes X_{i+1}X_{i+2}+
    \sum_{i=L}^{2L-1} \ketbra{i}\otimes Z_{i-L+1},
\end{equation}
which acts on $\mathcal{H}_a\otimes \mathcal{H}_s$. It is easy to check that $\bra{G}U_H\ket{G}=\frac{H}{L(1+g)}$. One can also check that the norm of $\frac{H}{L(1+|g|)}$ is smaller than $1$. In practice, the state $\ket{G}$ can be generated by a unitary operator $U_G$ which acts on the state $\ket{0}$ of the ancilla qubits as
\begin{equation}
  U_G\ket{0}_a = \ket{G}_a.  
\end{equation}
defining $\theta_g=2 \arccos{(1+g)^{-1/2}}$, one can check that the operator 
\begin{equation}
    U_G = RY(\theta_g)\otimes \left[\bigotimes_{i=1}^{n_a - 1}H\right],
\end{equation}
satisfies the aforementioned property. $RY(\theta)$ is the qubit rotation around the $Y$ axis and $H$ the Hadamard gate.

Since the Pauli $X$ and $Z$ both square to one, the unitary encoding satisfies the property $U_H^2 = I_{as}$. Therefore, following \cite{Low2019hamiltonian} and as explained in Sec. \ref{sec:qubitization}, the iterate $W_H$ is obtained by composing $U_H$ with a reflection w.r.t $\ket{G}$,
\begin{equation}
    W_H = \left[(2\ketbra{G}-I_a)\otimes I_s\right]U_H.
\end{equation}
The reflection $R_G=2\ketbra{G}-I_a$ can be implemented in the following way: since
\begin{equation}
    R_G = U_G(2\ketbra{0}-I_a)U_G^{\dagger},
\end{equation}
the only thing we need to construct is the reflection w.r.t. $\ket{0}$. It is easy to see that a multi-controlled $Z$ gate (or $C^nZ$ gate)\footnote{The multi-controlled version of a gate $U$, involving $n$ control qubits, will be denoted $C^nU$. The anti-multi-controlled version, i.e. the one that is executed only if all the control qubits are in the state $\ket{0}$, will be denoted $\widebar{C}^nU$} is an anti-reflection w.r.t. the state where all the qubits are in the state $\ket{1}$. Therefore an anti-reflection w.r.t. the state $\ket{0}_a$ is obtained by conjugating $\widebar{C}^nZ$ with the operator $\bigotimes_{i=1}^{n_a}X$,
\begin{equation}
    -R_G = U_G \left[\bigotimes_{i=1}^{n_a}X\right]\, \widebar{C}^nZ \left[\bigotimes_{i=1}^{n_a}X\right] U_G^{\dagger}.
\end{equation}
We are interested in the anti-reflection since ultimately the operator we want to construct is $-W_H$, see Eq. \eqref{CircuitTotal}.

\paragraph{Total number of qubits and gates} 

The final operator we would like to construct involves $C^{n_{\widebar{a}}}W_H^k$ (see Eq. \eqref{CircuitTotal}). To construct it we first build a controlled version of $W_H^k$ with one control qubit $A_0$ (a clean ancilla qubit) and then use $C^{n_{\widebar{a}}}X$ gates to implement $C^{n_{\widebar{a}}}W_H^k$. In circuit form it corresponds to 
\begin{equation}
    \begin{quantikz}
     \lstick{$\widebar{a}$} &\qw   & \ctrl{1}  & \qw & \ctrl{1} &  \qw 
    \\
    \lstick{$\ket{0}_{A_0}$} & \qw  & \gate{C^{n_{\widebar{a}}}X} & \ctrl{1} &\gate{C^{n_{\widebar{a}}}X} & \qw 
    \\
    \lstick{$a,s$} & \qw & \qw & \gate{CW_H^k} & \qw & \qw
    \end{quantikz}
    \quad.
\end{equation}
The number of qubits to simulate the system's Hilbert space corresponds to the number of sites $L$. Moreover, $O(\log L)$ qubits are necessary for the unitary encoding of the Hamiltonian. Finally $O(\log N)$ qubits are necessary for the unitary encoding of the Chebyshev expansion, therefore the total number of qubits scales as 
\begin{equation}
    \#\mathrm{qubits} \sim L + \log N,
\end{equation}
when both $L$ and $N$ are large.

We can also estimate the number of Toffoli gates. In order to implement a $C^nX$ gate, $O(n)$ Toffoli gates are necessary. Now since the Hamiltonian contains $L$ sites and that the unitary encoding involves $O(\log L)$ qubits, we need to implement an $O(L)$ amount of $C^{\log L}X$ gates, which corresponds to $O(L\log L)$ Toffoli gates for the unitary encoding of $H$. The addition of the reflection to construct the iterate does not change the scaling. For the total circuit, we need to implement all the powers of the (controlled) iterate until the $N$th power. The $k$th power of the (controlled) iterate involves $O(k L \log L)$ Toffoli gates. For the construction of the $C^{n_{\widebar{a}}}W_H^k$ we also need to use two $C^{n_{\widebar{a}}}X$ gates, which require $O(\log N)$ additional Toffoli gates. The total number of Toffoli for the implementation of $C^{n_{\widebar{a}}}W_H^k$ is therefore $O(k L \log L + \log N)$. Moreover the algorithm applies successively all the powers until $k=N$, therefore\footnote{Since $\sum_{k=1}^N( k L \log L + \log N)= O(N^2 L \log L + N\log N)$.}
\begin{equation}
    \#\mathrm{Toffoli} \sim N^2 L\log L+ N\log N.
    \label{Toffolis}
\end{equation}

\begin{figure}[htbp]
\centering
\includegraphics[width=.4\textwidth]{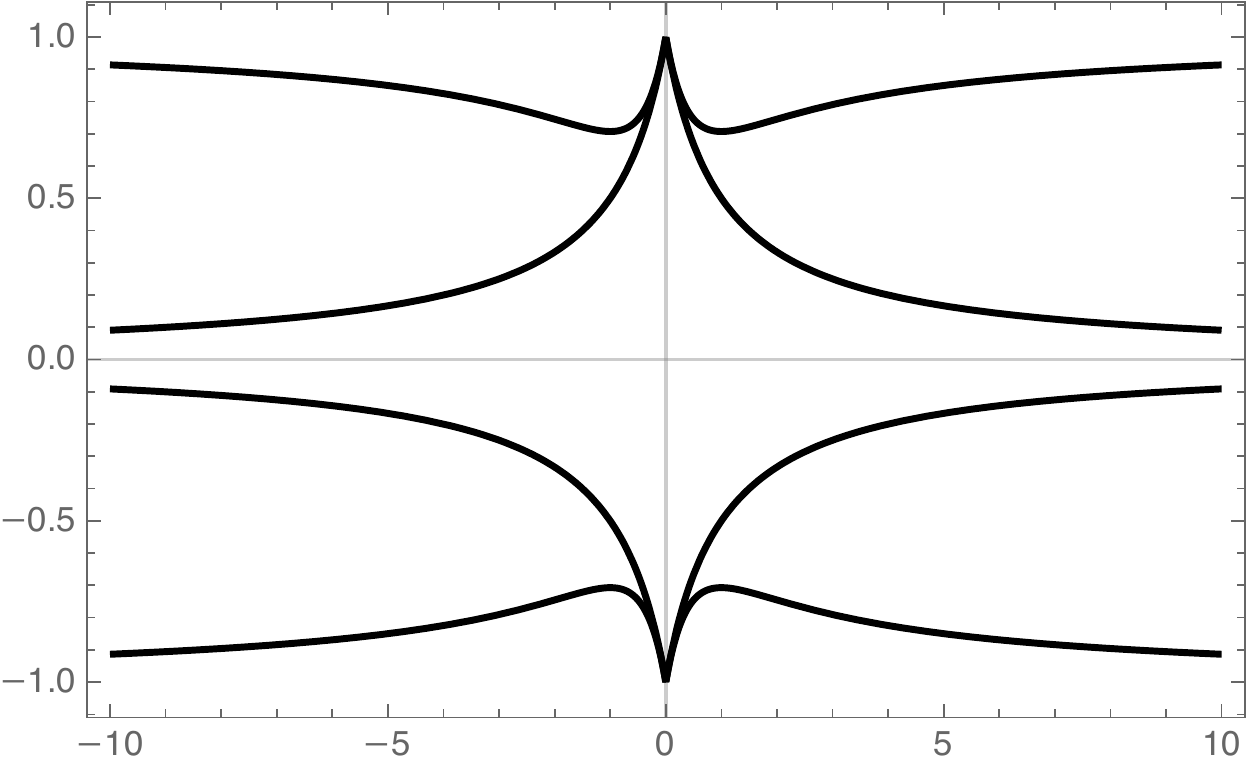}
\qquad
\includegraphics[width=.54\textwidth]{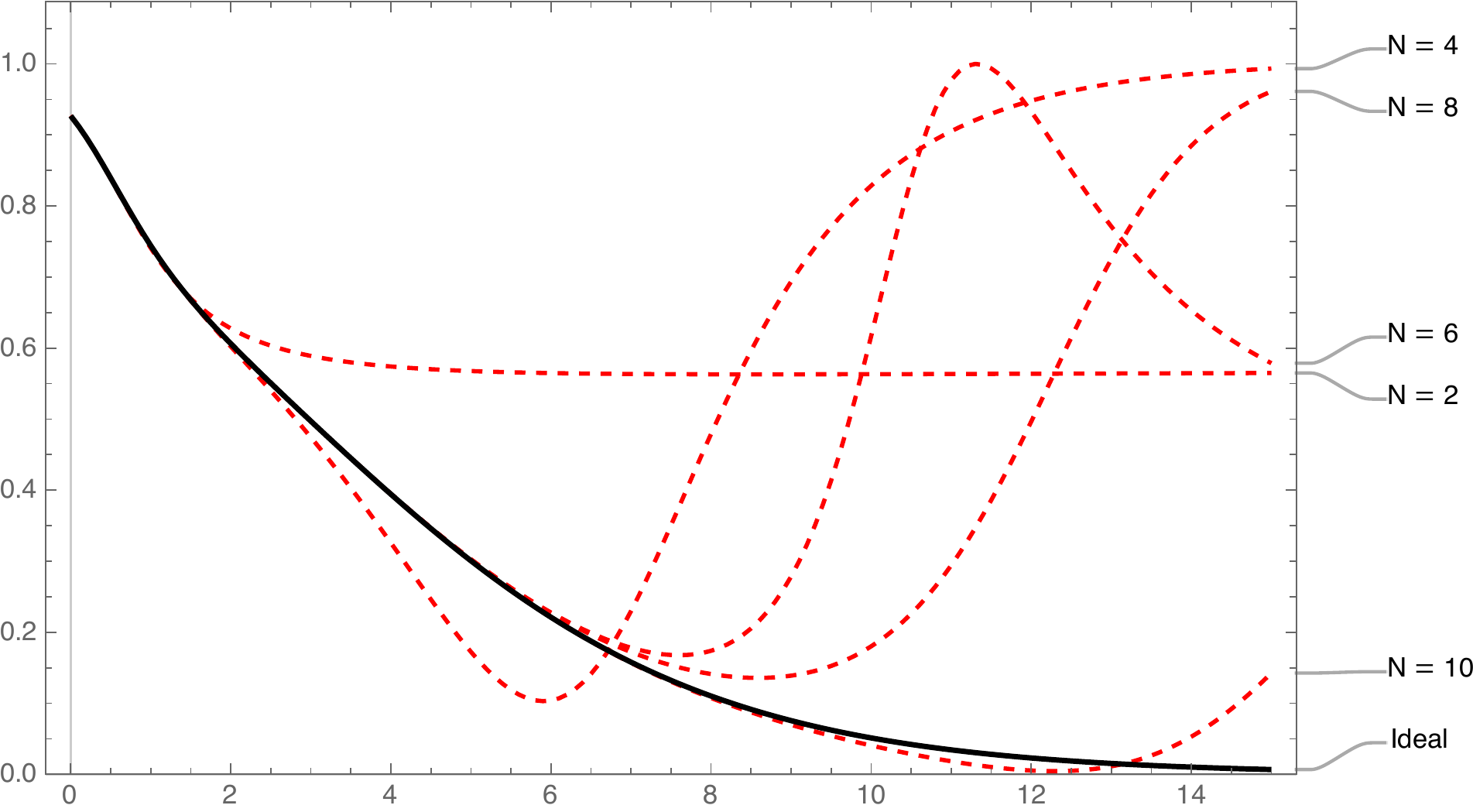}
\caption{\emph{Left}: Eigenvalues of the $L=2$ encoded Hamiltonian. We observe that the absolute value of each eigenvalue is bounded by one.
\emph{Right}: Numerical plot of the theoretical distance -- i.e. without any physical or simulated noise -- between the truncated Euclidean evolution of the state $\ket{\psi} = \frac{1}{\sqrt{2}}\left(\ket{00}_s+\ket{01}_s\right)$ (whose overlap with the first excited state is non-trivial) and the ground state as a function of the Euclidean time $\tau$. The parameters of this plot are $g=1$ and $N = \{2, 4, 6, 8, 10\}$. We have also plotted the ideal result. We observe the expected decrease of the distance followed by a growth due to the truncation of the Chebyshev expansion. One can also compute the lower bounds $\bar{\tau}_*$ (see Eq. \eqref{DistanceBound}) associated to each one of these curves: $\{(N, \bar{\tau}_*)\} = \{(2, 1.80),(4, 3.43), (6, 5.01), (8, 6.58), (10, 8.12)\}$. The bounds are consistently smaller than the actual crossover time $\tau_*$ of each curve. The corresponding upper bounds on the minimal distance are $\{(N, D^{\mathrm{min}}_{\mathrm{bound}})\} = \{(2, 3.24),(4, 1.65), (6, 0.85), (8, 0.45), (10, 0.24)\}$ where we observe that the bound becomes relevant when $N$ is sufficiently large. The $N=4$ and $N=10$ curves are interesting as in these cases the effect of the truncation error is first to amplify the decrease of the distance as compared to the ideal distance.}
\label{fig:eigenvalues}
\end{figure}

\paragraph{\texorpdfstring{$L=2$}{L=2} case} To illustrate our study of the TFIM we consider the two-sites case. When $L=2$ the dimension of the Hamiltonian is $4\times 4$ and the eigenvalues are 
\begin{equation}
-\frac{\sqrt{1+g^2}}{1+|g|}, \quad -\frac{1}{1+|g|}, \quad \frac{1}{1+|g|}, \quad \frac{\sqrt{1+g^2}}{1+|g|},
\end{equation}
see Fig. \ref{fig:eigenvalues}.

The first one being the energy of the following ground state
\begin{equation}
    \ket{\Omega} = \frac{\left(g - \sqrt{1+g^2} \right)\ket{00}_s + \ket{11}_s}{\sqrt{1+\left(g - \sqrt{1+g^2} \right)^2}}
\end{equation}
The second eigenvalue gives the energy of the first excited state
\begin{equation}
    \ket{\mathrm{exc}} = \frac{1}{\sqrt{2}}\left(\ket{10}_s-\ket{01}_s\right),
\end{equation}
which interestingly does not depend of the coupling $g$. 

As an example, running the $L=2$ algorithm with a truncation order $N = 10$ would require a total of $8$ qubits and, according to Fig. \ref{fig:eigenvalues}, the prepared state would be a very good approximation of the ground state. Current quantum computers can support up to a hundred noisy qubits which means that the number of qubits is not an issue. The bottleneck is the number of gates required to implement the algorithm. According to Eq. \eqref{Toffolis}, the number of Toffoli gates should be of order $10^2$, which in native gates would become $O(10^3)$. With this amount of native gates, any meaningful signal will be drowned in noise, which makes the application of our algorithm to the $L=2$ TFIM virtually impossible to implement on a currently existing quantum computer.

\section{A toy model experiment}
\label{sec:TOY}

Since the TFIM leads to a circuit with far too much depth to be implemented on a realistic quantum computer, we would like to look for the ground state of a simpler model. For this purpose we will consider a system whose Hamiltonian is 
\begin{equation}
    H=X.
\end{equation}
The eigenvalues are $\pm 1$ and the eigenstate are $\frac{1}{\sqrt{2}}(\ket{0}\pm\ket{1})$. An obvious unitary encoding is obtained by projecting the $CX$ gate on the $\ket{1}_a$ state. Therefore we have $\mathrm{dim}\mathcal{H}_s = \mathrm{dim}\mathcal{H}_a =1$. Moreover we will consider a Chebyshev expansion with truncation order $N=2$, which requires the dimension of  $\mathcal{H}_{\widebar{a}}$ to be $\mathrm{dim}\mathcal{H}_{\widebar{a}} = 4$. Finally we will need one additional ancilla qubit for the implementation of the $C^3X$ gates. The total number of qubits is therefore $n_{\mathrm{tot}}=5$. 

The unitary encoding $CX$ satisfies $CX^2= I$, which means that the iterate is obtained by multiplying the $CX$ gate by a reflection w.r.t. the state $\ket{1}_a$,
\begin{equation}
    W_X =\left[ (2\ketbra{1}-I_a)\otimes I_s\right]CX = -(Z\otimes I_s)CX.
    \label{WX}
\end{equation}

\begin{figure}[htbp]
\centering
\includegraphics[width=.9\textwidth]{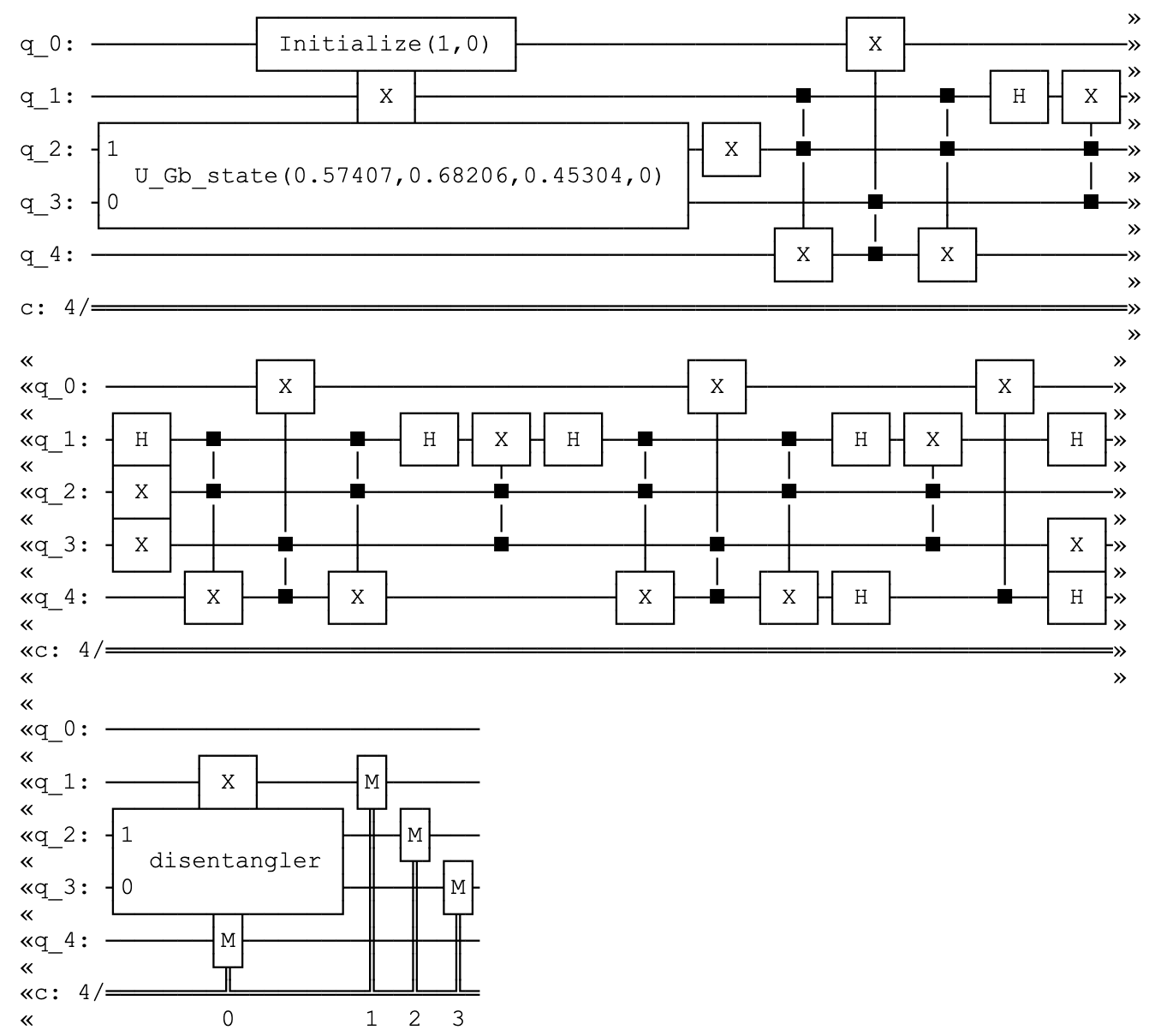}
\caption{Circuit for the preparation of a $\tau = 2.05$, $N=2$ truncated Euclidean evolution of the state $\ket{0}_s$. The qubit $q_0$ is the system qubit, the qubit $q_1$ is the ancilla-$a$ qubit, $q_2$ and $q_3$ are the ancilla-$\bar{a}$ qubits and $q_4$ is the ancilla-$A$ qubit. The circuit starts by initializing the system, ancilla-$a$ and ancilla-$\bar{a}$ qubits in the states $\ket{0}_s$, $\ket{1}_a$ and $\ket{\bar{G}}_{\bar{a}}$ (see Eq. \eqref{GbarState}) respectively. Then it implements the operator $\sum_{n=0}^2 \ketbra{n}_{\bar{a}}\otimes (-W_X)^n$ (see Eq. \eqref{WX}). The circuit ends with an $X$-gate to bring back $q_1$ to the state $\ket{0}_a$ and a disentangler whose role is to uncompute the state of the qubits 2 and 3 and bring them to the state $\ket{00}_{\bar{a}}$, followed by a measurement of ancilla-$a$ and ancilla-$\bar{a}$ qubits. The last $CX$ gate channels the measure of energy -- or equivalently the distance between the system state and the ground state $\frac{1}{\sqrt{2}}(\ket{0}_s-\ket{1}_s)$ -- to the ancilla-$A$ qubit which is also measured.}
\label{CircuitToy}
\end{figure}

The outcome of our algorithm will be a state which, when projected on $\ket{0000}_{Aa\widebar{a}}$ should give the $N=2$ truncated Euclidean evolution of some initial state $\ket{\psi}_s$:
\begin{equation}
    \ket{0}_A\left(\ket{000}_{a\widebar{a}}\ket{\widetilde{\psi}(\tau)}_s+\ket{\perp}_{a\widebar{a}s}\right),
    \label{StateToy}
\end{equation}
where $\ket{\perp}_{a\widebar{a}s}$ is orthogonal to $\ket{000}_{a\widebar{a}}$ and $\ket{\widetilde{\psi}(\tau)}_s$ is the un-normalized version of the state Eq. \eqref{TruncatedEvolvedState}. On top of the main circuit we would like to add a piece of circuit dedicated to the measure of the energy of the system. Since our Hamiltonian's energies are $+1$ or $-1$ we can channel this measure to the ancilla qubit $\mathcal{H}_A$ by applying the gates $H_A (C_A X_s) H_A$ to the state in Eq. \eqref{StateToy}. One obtains 
\begin{equation}
    \ket{000}_{a\widebar{a}}\left(\braket{\Omega}{\widetilde{\psi}(\tau)} \ket{1}_A\ket{\Omega}_s+\braket{\mathrm{exc}}{\widetilde{\psi}(\tau)} \ket{0}_A\ket{\mathrm{exc}}_s\right)+\ket{\widetilde{\perp}}_{Aa\widebar{a}},
\end{equation}
where $\ket{\widetilde{\perp}}_{Aa\widebar{a}}$ is another state perpendicular to $\ket{000}_{a\widebar{a}}$. We conclude that the distance between $\ket{\psi(\tau)}$ and the ground state is given by 
\begin{equation}
    D_{\Omega,\psi(\tau)} \equiv 1 - |\braket{\Omega}{\psi(\tau)}|^2 = p_\tau\left[\ket{0000}_{a\widebar{a}A}\right],
\end{equation}
where $p_\tau\left[\ket{0000}_{a\widebar{a}A}\right]$ is the probability that the outcome of a measurement of all the ancilla qubits gives $\ket{0000}_{a\widebar{a}A}$, provided we only consider the two outcomes $\ket{0000}_{a\widebar{a}A}$ and $\ket{0001}_{a\widebar{a}A}$. The circuit that implements the algorithm followed by the measurement of this probability is depicted in Fig. \ref{CircuitToy}.

\begin{figure}[htbp]
\centering
\includegraphics[width=.6\textwidth]{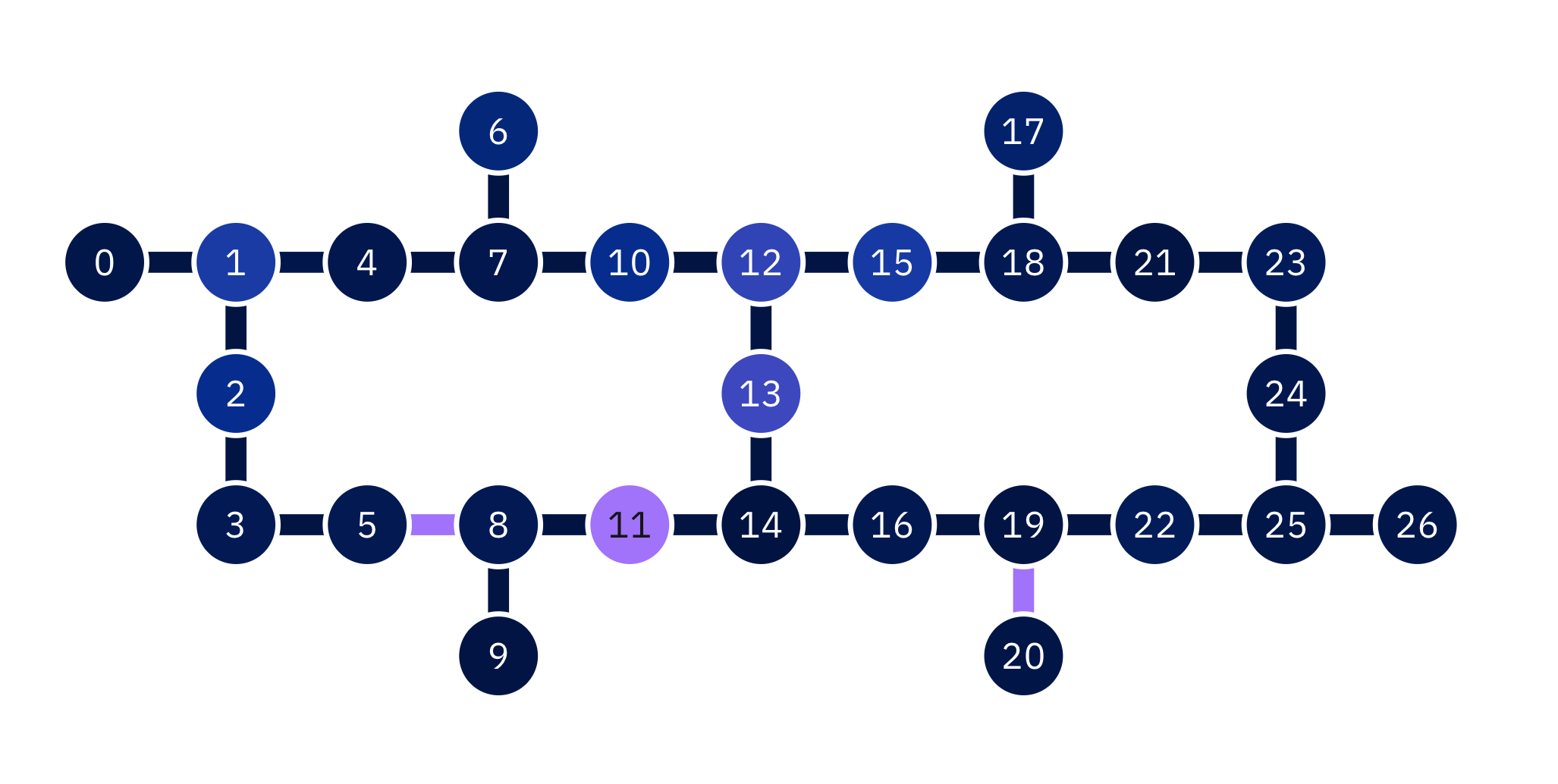}
\caption{Architecture of the $\texttt{ibm\_hanoi}$ 27 qubits processor. The pairs of qubits that are linked are pairs on which $CX$ gates can be performed.}
\label{fig:architecture}
\end{figure}

The experiment will be carried on Qiskit, a python package dedicated to quantum circuit. The latter allows to run a circuit with the addition a simulated noise that mimics the real noise of one of IBM's quantum processors. The one we choose to simulate is called $\texttt{ibm\_hanoi}$, it is a 27 qubits processor whose architecture is displayed in Fig. \ref{fig:architecture}.

\begin{figure}[htbp]
\centering
\includegraphics[width=0.8\textwidth]{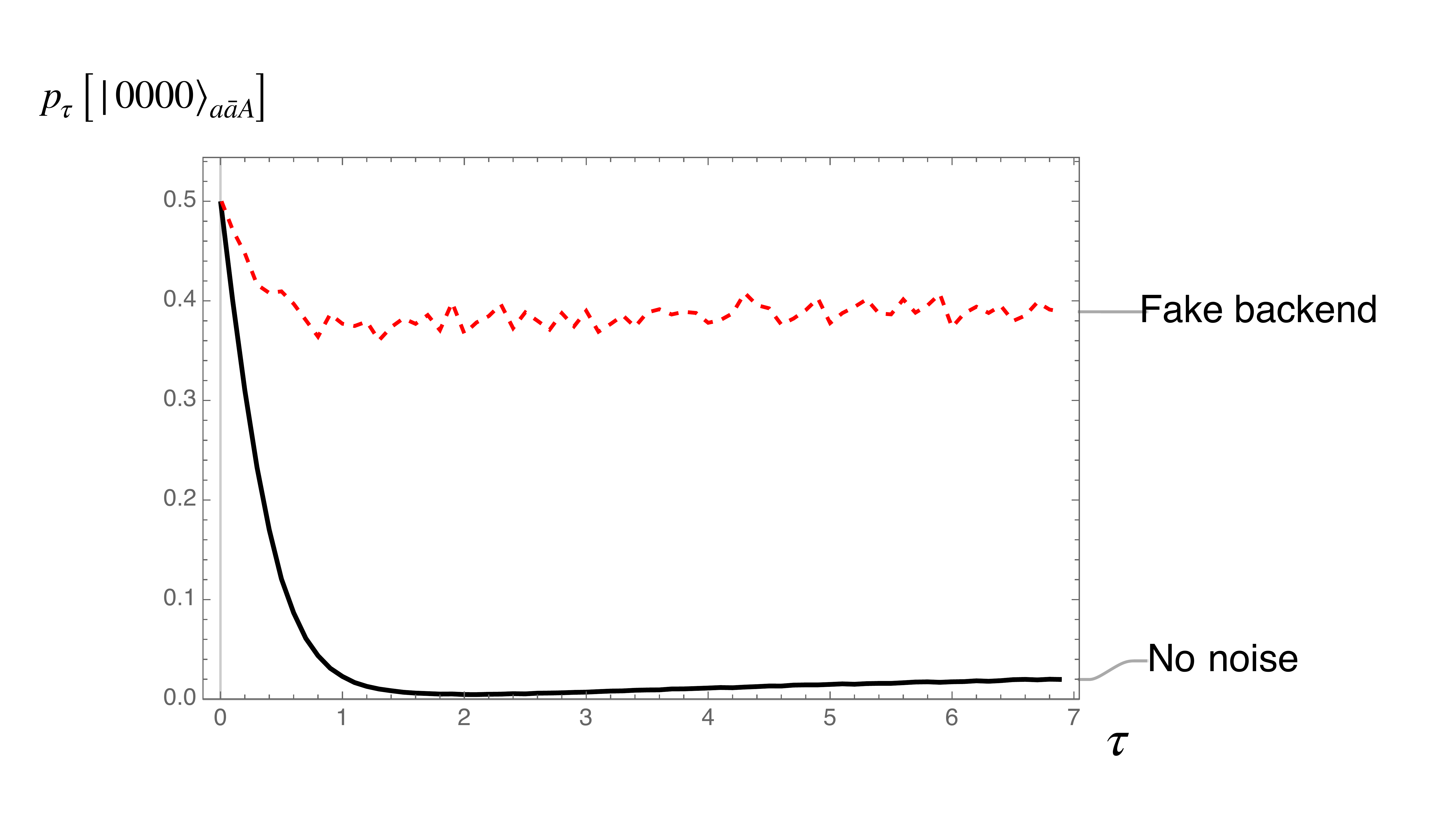}
\caption{Plot of the distance or equivalently $p_\tau\left[\ket{0000}_{a\bar{a}A}\right]$ as a function of the time $\tau$. The solid curve corresponds to an execution of the algorithm without any noise while the dashed curve is the same experiment but with the addition of the simulated noise of the real backend $\texttt{ibm\_hanoi}$. For both experiments we choose a time step $d\tau = 0.1$, at each time step we run the algorithm $n_{\mathrm{runs}} = 20$ times and each run involves a sampling of $p_\tau\left[\ket{0000}_{a\bar{a}A}\right]$ with $n_{\mathrm{shots}} = 2^4$ shots. We then average the probability obtained at each time over the $n_{\mathrm{run}} = 20$ runs and plot the result. We observe the expected behaviour, the distance starts with an exponential decrease and then increases again after a critical time $\tau_* = 2.05$. The signal with the simulated noise is considerably less sharp but interestingly it is not completely washed away by the noise and exhibits the same qualitative behaviour. We haven't added any layer of error mitigation to the experiment since this is not the subject of this paper but one could consider using zero noise extrapolation to obtain a signal closer to the non noisy one, at the usual  cost of running more circuits.}
\label{fig:toymodel}
\end{figure}

The code for the experiment is available in the following \href{https://github.com/Charles-Marteau/Ground-State-Preparation-via-Qubitization}{github repository} and the results are given in Fig. \ref{fig:toymodel}. The simulated noise of the fake backend is based on a snapshot of the characteristic of the real $\texttt{ibm\_hanoi}$ backend, such as coupling map, basis gates, qubit properties (T1, T2, error rate, etc).\footnote{See Qiskit documentation for more details on fake backends: \url{https://qiskit.org/documentation/apidoc/providers_fake_provider.html}.} All this information about the fake backend can be found in the code.

\newpage

\section*{Acknowledgments}

We thank Aidan Chatwin-Davies and David Wakeham for insightful discussions and comments on the draft, we also thank the String Theory Group of UBC and acknowledge support from NSERC.

\bibliography{biblio.bib}
\bibliographystyle{JHEP}

\end{document}